# Post-spontaneous-symmetry-breaking power-laws after a very strong earthquake: Indication for the preparation of a new strong earthquake or not?

Stelios M. Potirakis [1,*], Yiannis Contoyiannis [1,2], Konstantinos Eftaxias [2], Nikolaos S. Melis [3], and Constantinos Nomicos [1]

[1] Department of Electrical and Electronics Engineering, University of West Attica, Ancient Olive Grove Campus, 250 Thivon and P. Ralli, Aigaleo, Athens GR-12244, Greece; yiaconto@uniwa.gr (Y.C.), cnomicos@uniwa.gr (C.N.).
[2] Department of Physics, University of Athens, Panepistimiopolis, GR-15784, Zografos, GR-15784 Athens, Greece; ceftax@phys.uoa.gr.
[3] Institute of Geodynamics, National Observatory of Athens, Lofos Nimfon, Thissio, GR-11810 Athens, Greece; nmelis@noa.gr.
* Correspondence: spoti@uniwa.gr; Tel.: +30-2105381550.

**Abstract**: It has recently been found that the evolution of the preparation of a strong earthquake (EQ), as it is monitored through fracture-induced electromagnetic emissions (EME) in the MHz band, presents striking similarity with the evolution of a thermal system as temperature drops, since distinct steps of the evolution of the phenomenon of spontaneous symmetry breaking (SSB) can be identified. Here, the study of fracture-induced EME in the MHz band in analogy to thermal systems is extended to the phase of local fracture structures that follow after the SSB (and the occurrence of the main EQ). By comparing fracture-induced MHz EME associated with the strongest EQs ($M_W = 6.9$) that occurred in Greece during the last twenty years with the 3D Ising model, a way to distinguish whether a possible identification of post-SSB power-laws immediately after a very strong EQ is a sign for the preparation of a new strong EQ or not is provided. In the suggested approach, the time series analysis method known as the method of critical fluctuations is used, enhanced by the autocorrelation function, while the role of the latter proves to be decisive.

## 1. Introduction

The field of phase transition phenomena is an important field of statistical Physics that has extensively been studied for thermal systems but has also found applications to many different disciplines. Phase transition refers to the transition between two phases (states) in which a system could exist. It has been proposed, e.g., [1], that as the lithosphere system evolves towards an intense earthquake (EQ) it can be studied from the phase transitions point of view. This suggestion has been investigated in various ways within the framework of a phase transition that occurs during the materials' rapture, starting from the level of the microcracking dynamics and reaching the percolation process across the whole material, e.g., [2, 3]. On the basis of the collective phenomena occurring in critical systems, it has been suggested that intense EQs result after the lithosphere has reached a kind of critical point (e.g., [4] and references therein). A manifestation of critical state in a system experiencing a phase transition is the appearance of power-laws in its observables. Such power-law findings have repeatedly been reported for fracture phenomena (e.g., [5-7]) implying scaling behaviors.

A research effort to study EQ preparation in analogy to thermal systems through the analysis of fracture-induced electromagnetic emissions (EME) in the MHz band has been continuing since more than 15 years, e.g., [8-13]. This effort builds on the recordings of ELSEM-Net (HELlenic Seismo-ElectroMagnetics Network) (http://elsem-net.uniwa.gr, see also supplementary downloadable material of [13] and [14]), our ground-based network of telemetric stations spanning all over Greece, continuously recording MHz – kHz EME at a sampling period of 1 s. From [10-14] the following conclusion can be drawn: A few days or even a few hours before a main EQ event that happens on land or near coastline, has magnitude > 5.5 and relatively shallow hypocenter, MHz EME time series excerpts, termed "critical windows", are recorded, the fluctuations of which are proven critical in the terms of the Physics of critical phenomena. This criticality in MHz EME manifests itself through the existence of power-laws in the distribution of properly defined waiting times known as "laminar lengths" and can be identified by means of the time series analysis method known as the method of critical fluctuations (MCF) [13, 15]. In all cases of strong EQ events which have occurred on-land or near coastline in Greece and have been studied up to now [12, 13] we have found the above-mentioned critical windows before the EQ event. Unfortunately, the inverse is not always valid. That is, finding MHz EME critical windows does not mean that a strong EQ event will always follow [10, 12].

Recently, attempting to gain a deeper understanding to the involved EQ preparation processes and the organization to critical state in terms of the mechanism of Lévy flight and its possible engagement with Gaussian stochastic processes, it was found that strong events occur when the exponent values of the power-law for laminar lengths distribution is limited within the interval (1-1.5) [12]. In the specific work, further to the MCF analysis, the application of the autocorrelation function to pre-EQ MHz EME was also employed to discriminate between criticality associated with the preparation of strong EQ events and that associated with swarms of weaker EQs, since for the second case a ''collapse'' of autocorrelation is observed on the longer scales. Moreover, it was recently shown that the phenomenon of spontaneous symmetry breaking (SSB), that appears during a second order phase transition from the symmetric phase (critical state) to a low symmetry phase (broken symmetry phase), can be monitored in real-time through the analysis of MHz EME in the form of distinct steps along the preparation of a strong seismic event [11]. Specifically, a model through which one can study the SSB, the Ising model in 3 dimensions (3D Ising), has been employed in [11]. In the specific work it has been shown (Fig. 1 in [11]) that the critical point is degenerated into a narrow temperature zone, where the breaking of the symmetry occurs gradually as the temperature drops. The evolution of the SSB inside the zone, according to [11, 16] has as follows: At critical point the effective potential has a vacuum at zero where the mean magnetization is zero ($M = 0$). As the symmetry begins to break two new vacua appear. These vacua communicate with each other and as the temperature drops they gradually separate. When the separation of the vacua is completed, then the SSB is completed. The narrow zone between the critical temperature and the temperature where the SSB is completed is hereafter referred to as the "critical zone". After this the system selects one of the two vacua and the completion of SSB has been accomplished.

In the present work we extend our investigation to the time period after a strong EQ event, based on the physical context of broken symmetry phase after the SSB phenomenon. For this purpose, we compare two very strong ($M_W = 6.9$) EQs that occurred in Greece with the 3D Ising model which obeys critical behavior, providing a criterion to whether post-SSB power-laws after a very strong EQ can be considered an indication for the preparation of a new strong EQ or not. The time series analysis is performed by means of the MCF, enhanced by the calculation of the autocorrelation function. Monitoring of the autocorrelation function values for time series excerpts that present power-law behavior after strong EQ events is proved to be of key importance.

**2. Power-laws surviving after the spontaneous symmetry breaking**

For a Z(N) spin system, spin variables are defined as: $s(a_i) = e^{i2\pi a_i/N}$ (lattice vertices $i = 1 \ldots i_{max}$), with $a_i = 0,1,2,3 \ldots N-1$. Specifically, for $N = 2$ and for 3 dimensions we consider the 3D Ising model. An effective algorithm which produces configurations for the 3D Ising model is the Metropolis algorithm. In this algorithm the configurations at constant temperatures are selected with Boltzmann statistical weights $e^{-\beta H}$, where $H$ the Hamiltonian of the spin system with nearest neighbors' interactions, which can be written as:

$$H = -\sum_{<i,j>} J_{ij} s_i s_j \tag{1}$$

It is known [16] that this model undergoes a second-order phase transition when the temperature drops below a critical value. Thus, for a $20^3$ lattice the critical temperature has been found to be $T_c = 4.545$ (for $J_{ij} = 1$), while the end of the critical zone, when the SSB phenomenon is completed, occurs for $T_{SSB} = 4.44$ [11]. In every temperature within the critical zone the dynamics of the fluctuations of magnetization is the intermittency type I with $M = 0$ as a marginal unstable fixed point [15]. This dynamics is expressed through power-laws in the distribution of laminar lengths (i.e., waiting times within the laminar region) $L$ [17] as:

$$P(L) \sim L^{-p} \tag{2}$$

In the following we present the results obtained for a numerical experiment of a $20^3$ lattice 3D Ising that has been performed for 100000 iterations. The algorithmic time $\tau$ corresponds to sweeps of the lattice. So, a "time" series $M(\tau)$ is produced. In Fig. 1 we present the distributions of laminar lengths, calculated by MCF [13, 15], for two temperatures: one for the critical temperature $T_c$ and one more for $T = 4.4$ for which the symmetry has been broken but is still very close to the $T_{SSB}$. The fit on these distributions is accomplished using the function:

$$f(L) \sim L^{-p} e^{-qL}, \tag{3}$$

where the exponential factor is added for the description of finite-size effects. The MCF provides the criteria of criticality in terms of the values of the exponents $p, q$.

Specifically, it must be $p \in (1,2)$ and $q$ very close to zero [11-13]. In the ideal case it is $q = 0$. Due to the fact that the temperature $T = 4.4$ ($<T_{SSB}$), although lower than $T_{SSB}$, it is still very close to $T_{SSB}$, it could be expected that the criticality criteria could still be (marginally) satisfied. But in such a case the value of $p$, that is the exponent of the power-law factor, should lie close to the lower limit for criticality which is $p = 1$. A smaller temperature than $T = 4.4$ would no longer give any indications of critical state.

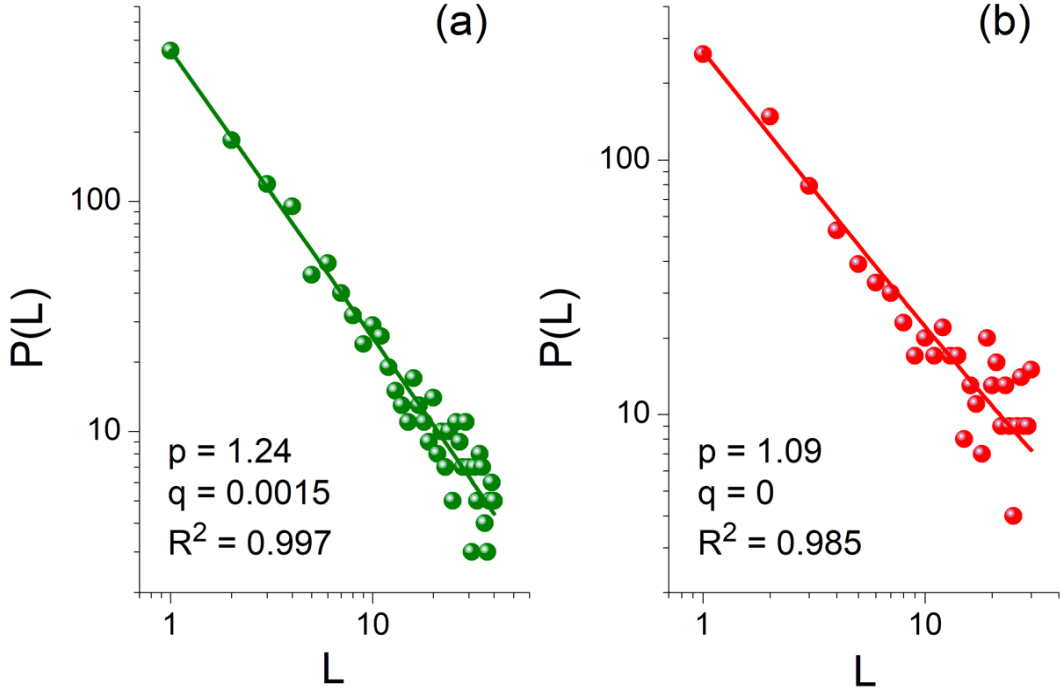

**Fig. 1.** (a) The laminar lengths distribution at critical temperature $T_c = 4.545$ for the numerical experiment described in the text. The exponents obtained for the fitting function of Eq. (3) are $p = 1.24$ and $q = 0.0015$, with a goodness of fit $R^2 = 0.997$. These values are compatible to a power-law in critical state. (b) The laminar lengths distribution at temperature $T = 4.4$ ($<T_{SSB}$) very close to the critical zone. The exponents obtained for the fitting function of Eq. (3) are $p = 1.09$ and $q = 0$, with a goodness of fit $R^2 = 0.985$. Note that as temperature drops from $T_c = 4.545$ to $T_{SSB} = 4.44$, the SSB phenomenon is completed as evident by the order parameter values' distribution shape change until the appearance of two separated lobes when the two degenerated vacua completely separate (Fig. 1 of [11]).

The results demonstrated in Fig. 1b verify that after SSB but very close to it the power-law in the laminar lengths distribution, implying criticality, "survives" although the unstable critical point no longer exists. As the temperature drops to values lower than $T = 4.4$ no indications of criticality can be found anymore. In the Ising model after SSB a phase transition is accomplished. From the symmetry phase, where the mean magnetization is zero, the system goes to the broken symmetry phase, where the mean magnetization is not zero anymore. Up to the point that SSB is completed the system lies in the critical state. An important characteristic of the critical state is that the correlations between the system parts dominate in all scales. That is manifested by the fact that the autocorrelation function presents long memory [16] at the critical state. On the contrary, as the temperature drops below $T_{SSB}$ the autocorrelation function values collapse very quickly. This is due to the creation of local structures, known as domain walls [16], in the new phase. Domain walls do not permit the "communication" throughout the whole extension of the system and, accordingly, long correlation lengths do not appear anymore.

In the abovementioned framework, the autocorrelation function for $T_c = 4.545$ and $T = 4.4$ ($<T_{SSB}$) (Fig. 2), provides clear evidence discriminating the nature of the power-laws of Fig. 1. The power-law at $T_c = 4.545$ corresponds to critical state, while the power-law at $T = 4.4$ ($<T_{SSB}$) is a result of "locally surviving" critical dynamics although the unstable critical point no longer exists. Thus, any confusion that might be caused by the existence of power-law close after the completion of SSB is restored by means of the autocorrelation function, clearly showing the collapse of long-range correlation during the phase of the domain walls.

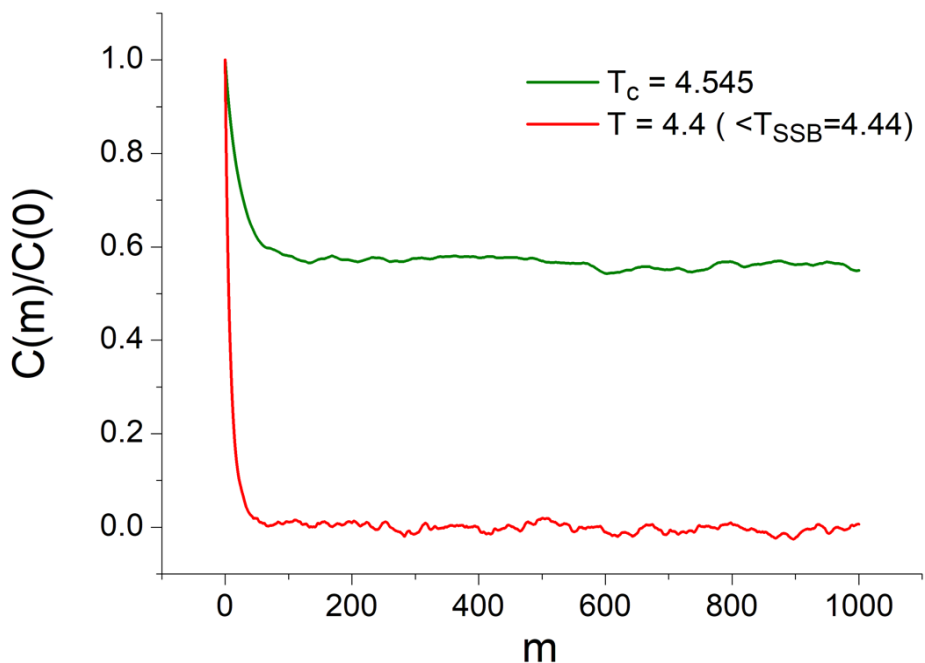

**Fig. 2.** The normalized autocorrelation function vs. correlation length produced from the "time" series of numerical experiment at critical temperature (green curve) and at $T = 4.4$ (red curve).

As it is shown in Sec. 3, such phenomena appear after the creation of very strong EQs. This is because in the case of very strong (high magnitude) seismic events, due to the longer distances of propagation of the correlations at the critical state, the autocorrelation function is expected to retain a longer memory in the critical state and thus autocorrelation collapse phenomena are easier to distinguish.

**3. Investigating power-laws appearing after strong earthquakes**

On 30-10-2020, a very strong EQ took place off the coast North of Samos Island (Greece): $M_W = 6.9$, epicenter (37.9001ºN, 26.8057ºE), focal depth = 12 km, time of occurrence 11:51:57 UTC [18, 19], hereafter referred to as "the 2020 Samos EQ". Pre-seismic MHz EME were recorded by the ELSEM-Net station located on Lesvos Island, which is relatively close to Samos (< 160 km away North from the EQ epicenter), two days before the EQ event. Specifically, a critical window of 14000 s width was recorded, the distribution of the laminar lengths of which by means of MCF is shown in Fig. 3a. Moreover, in analyzing the recordings of Lesvos Island in the MHz band using MCF, power-laws in the distribution of laminar lengths were identified again during the next two days after the EQ, for specific MHz recordings excerpts, as shown in Figs. 3b and 3c. Note that SSB was also identified, as described in detail in [11], in the MHz EME recordings of 29/10/2020, (not shown here), i.e., after the critical window of Fig. 3a and one day before the 2020 Samos EQ occurrence.

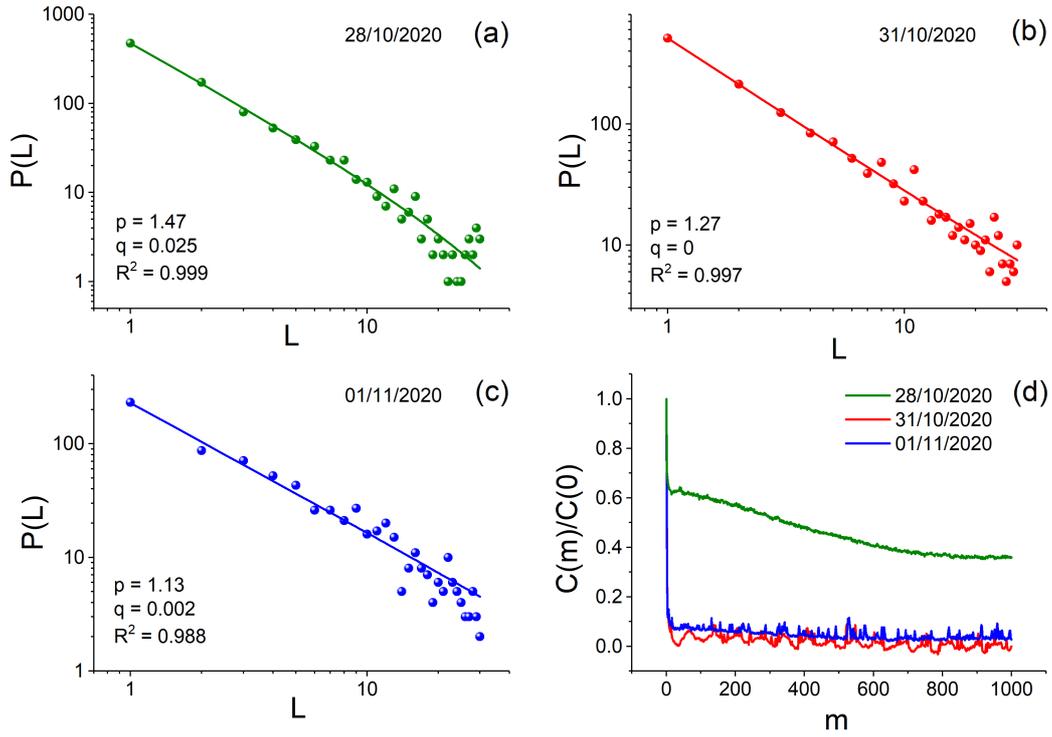

**Fig. 3.** Distribution of laminar lengths for the MHz time series excerpts corresponding to the recordings at Lesvos Island station within the time windows (figure format follows Fig. 1): (a) 33000-47000 s (i.e., 09:10:00-13:03:20 UTC) of 28/10/2020, two days before the 2020 Samos EQ; (b) 10000-27700 s (i.e., 02:46:40-7:41:40 UTC) of 31/10/2020, one day after the 2020 Samos EQ; (c) 53000-73000 s (i.e., 14:43:20-20:16:40 UTC) of 01/11/2020, two days after the 2020 Samos EQ. (d) The normalized autocorrelation function vs. correlation length produced from the MHz time series excerpts corresponding to the power-laws presented in Fig. 3a-c: green curve (Fig. 3a), red curve (Fig. 3b), blue curve (Fig. 3c).

If one overlooks the obvious drop of the value of the power-law exponent $p$ in the temporal sequence of the power-laws of Fig. 3a-c there is nothing else to differentiate them. Thus, since the up to now criteria for the identification of critical state according to MCF, i.e., $p \in (1,2)$ and $q$ very close to zero for the distribution of laminar lengths as fitted by the function of Eq. (3), are satisfied by all three of them, there is nothing to refraining one from concluding that all three of them indicate critical state and therefore are corresponding time series excerpts are "critical windows". Knowing that 2020 Samos EQ occurred after the criticality indicated by the power-law of Fig. 3a and its following SSB, it could be considered that the post-EQ sequence of power-laws of Figs. 3b and 3c (note also that $p < 1.5$ [12]) indicates the critical organization of a new strong EQ that has not yet occurred. However, as shown in Sec. 2, post-SSB power-laws in the laminar lengths' distribution may appear as a result of "locally surviving" critical dynamics although the unstable critical point no longer exists. Thus, in the up to now application of MCF to MHz EME, a criterion is missing that could answer the question whether the observed post-SSB power-laws after a very strong EQ are an indication for the preparation of a new strong EQ or not. This additional criterion, that would, beyond any doubt, verify whether an identified power-law in the laminar lengths of MHz EME is indeed a critical window that signalizes the "critical epoch" in the preparation of a new strong EQ, when the short-range correlations evolve into long-range ones, is the existence of long-range correlation expressed through persistently high values in the temporal evolution of the autocorrelation function values (cf. Sec. 2).

Indeed, the autocorrelation function (Fig. 3d) successfully discriminates the power-law of Fig. 3a from those of Figs. 3b,c. As in the case of the Ising model (Sec. 2), here too the autocorrelation function distinguishes the pre-EQ power-law that indicates critical state (Fig. 3a), and the two post-EQ power-laws (Figs. 3b, c) that correspond to the phase of the domain walls. According to [10, 11], in the process of SSB in a heterogeneous medium we discern two phases in analogy to the thermal phase transition. The first is the phase of microcracking organization in the heterogeneous system. This microcracking organization, as is has been shown in [12], follows a Lévy flight random walk mechanism. The randomly distributed directions of microcracking correspond to a symmetric phase analogous to the randomly distributed magnetic moments in a magnetic material at high temperatures. The final stage of this phase is the critical state where the long-range correlations are developed. The second phase comes after the SSB completion and corresponds to directional, localized, cracks where preferred directions of fracture structures have begun to form, in analogy to the phase of domain walls in magnetic systems. In the specific phase the large-scale correlations are collapsing.

As in thermal systems for which the control parameter is the temperature, for fracture phenomena it has been suggested that the control parameter that controls the distance to global rupture and can be time, strain, stress or combinations of these [3]. Although stress is usually considered as the control parameter both in the study of laboratory experiments and model studies [20, 21], time has also been highlighted as the control parameter of the failure process in laboratory and geophysical scale fracture phenomena [11, 22]. Following the chronological order for the 2020 Samos EQ case, we have that on 28/10/2020 the critical state around the focal area has been created and the long-range correlations have been developed (green curve in Fig. 3d). Following that, the EQ event took place on 30/10/2020, after the SSB phenomenon has been completed. In the next two days, 31/11/2020 and 01/11/2020, i.e., immediately after the symmetry has been broken, the appearance of post-EQ local fracture structures (domain walls) is the cause of the collapse of the correlations (red and blue curves in Fig. 3d), while critical dynamics "locally survive" due to the small temporal distance from the moment of the EQ occurrence (close after the completion of SSB).

Thus, the local fractures that occur immediately after strong EQs can give indications of locally surviving critical dynamics. However, this case is characterized by the collapse of the correlation function values revealing that this is not a critical organization. Therefore, in such a case, the observed post-SSB power-laws after a very strong EQ are not an indication for the preparation of a new strong EQ.

At this point, it is important to note that in the case of 2020 Samos EQ, as has been observed many times in the past [10], MHz EME embedding indications of critical dynamics (e.g., in the form of power-laws in the distribution of laminar lengths by means of MCF analysis) were not observed during the time period between the completion of SSB and the occurrence of the main EQ. This indicates that such post-EQ power-laws in the distribution of laminar lengths of MHz EME, that are not accompanied by long memory in the corresponding autocorrelation function, are not related to the main-shock preparation processes but are associated to local fractures in course of the aftershock sequence which are not able to organize the system towards the preparation of a new main shock.

It is also interesting to consider an example of a strong EQ that has been followed by another strong EQ a few days later. This would allow us to check whether any post-EQ power-laws in the distribution of laminar lengths of MHz EME are consistent with the use of autocorrelation function as a way to identify whether post-EQ power-laws are an indication for the preparation of a new strong EQ or not.

In 2008, a sequence of strong EQs took place in South-West Greece, offshore of Methoni town, thus referred to as the 2008 Methoni EQs. The specific EQ sequence consisted of three very strong EQs, the first two with magnitudes $M_W = 6.9$ and $M_W = 6.3$ occurred within a two-hour time interval on 14/02/2008, while a third one with magnitude $M_W = 6.1$ occurred on 20/02/2008, all shallow (< 30 km depth) and with neighboring epicenters (max inter-epicenter distance < 50 km), e.g. [23]. In Fig. 4 the distributions of the laminar lengths produced from three MHz EME excerpts are shown. These excerpts were recorded by the ELSEM-Net station located in Zakynthos (Zante) Island (~ 160 km away from the EQ epicenters) on 12/02/2008 (Fig. 4a), i.e., two days before the first one of the 2008 Methoni EQs, on 15/02/2008 (Fig. 4b) and on 16/02/2008 (Fig. 4c), i.e., after the first two EQs but before the third EQ of the specific sequence. In Fig 4d the autocorrelation functions of the corresponding MHz EME excerpts are presented. Following the chronological order for the 2008 Methoni EQs case, we have that on 12/02/2008 the critical state around the focal area has been created and the long-range correlations have been developed (green curve in Fig. 4d). Subsequently, the first two EQs took place on 14/02/2008, after the SSB phenomenon has been completed on 13/02/2008 (Fig. 2 of [11]). On 15/02/2008, i.e., immediately after the symmetry has been broken and the first two 2008 Methoni EQs have taken place, the appearance of post-EQ local fracture structures (domain walls) is the cause of the collapse of the correlations (red curve in Fig. 4d), while critical dynamics "locally survive" due to the small temporal distance from the moment of the EQ occurrence (close after the completion of SSB). So up to the specific time point, no indications for a new strong EQ exist. However, on the next day, 16/02/2008, the system is again in critical state, since long-range correlations have again been developed (blue curve in Fig. 4d), implying the preparation of a new strong EQ event.

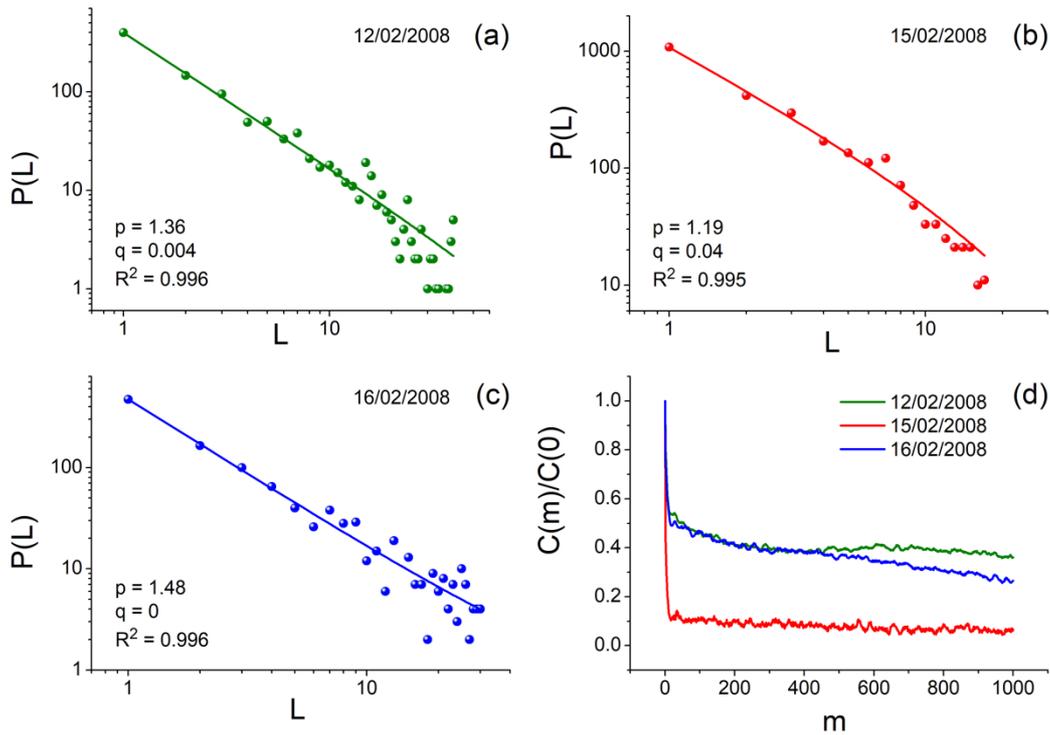

**Fig. 4.** Distribution of laminar lengths for the MHz time series excerpts corresponding to the recordings at Zakynthos Island station within the time windows (figure format follows Fig. 1): (a) 27000-46000 s (i.e., 07:30:00-12:46:40 UTC) of 12/02/2008, two days before the 2008 Methoni EQs; (b) 5000-25000s (i.e., 01:23:20-06:56:40 UTC) of 15/02/2008, one day after the first two 2008 Methoni EQs; (c) 43000-61000 s (i.e., 11:56:40-16:56:40 UTC) of 16/02/2008, two days after the first two 2008 Methoni EQs and four days before the third one of the 2008 Methoni EQs. (d) The normalized autocorrelation function vs. correlation length produced from the time series corresponding to the power-laws presented in Figs. 4a-c: green curve (Fig. 4a), red curve (Fig. 4b), blue curve (Fig. 4c).

Comparing Figs. 3 and 4 one can make the following remarks:

- The difference observed in Fig. 4d between the correlation length in the critical state (12/02/2008, green curve) and the correlation length of one day after the first two 2008 Methoni EQs (15/02/2008, red curve), indicates similar behavior as the one observed for the 2020 Samos EQ (see 28/10/2020 and 31/10/2020 curves in Fig. 3d), i.e., the autocorrelation function has collapsed immediately after the EQ.
- On the second day after the EQ the differentiation between Fig. 3d and Fig. 4d is characteristic. While in the case of 2020 Samos EQ the collapse of the autocorrelation function values persists on 01/11/2020 (Fig. 3d, blue curve), in the case of the 2008 Methoni EQs the autocorrelation function values on 16/02/2008 (Fig. 4d, blue curve)

- have almost returned to the level of the critical state before the first two 2008 Methoni EQs.
- By comparing the temporal sequence of the power-laws of Figs. 3a-c and Figs. 4a-c, one observes that for the 2020 Samos EQ the temporal evolution of the value of the power-law exponent $p$ is a monotonic drop towards the value that corresponds to the lower limit for criticality, $p = 1$, while for the case of the 2008 Methoni EQs this monotonic behavior is not present. Specifically, on the second day after the first two 2008 Methoni EQs (Fig. 4c) the exponent $p$ value is increased.

The above results are completely compatible with the evolution of each one of the two examined cases. In the case of 2020 Samos EQ, up to three months after the main shock, no strong EQ (M > 5.5) has been recorded in Greece with its epicenter close to the 2020 Samos EQ epicenter. So, its evolution was a sequence of small aftershocks. On the contrary, 6 days after the two first Methoni EQs, one more strong EQ occurred very close to them. So, one can conclude that the power-law of Fig. 4c refers to the critical organization of the new strong EQ that took place four days after the recording of the corresponding MHz EME time series excerpt. From the above-mentioned, the importance of the information that the autocorrelation function provides during the phase of domain walls, i.e., after the occurrence of a strong seismic event, is highlighted. In case that post-SSB power-laws are identified in the MHz EME recordings after a very strong EQ, then two alternatives are possible: if the corresponding autocorrelation function values collapse immediately after the EQ and remain low, then no new strong is expected, but if the autocorrelation function values return to high values then a new strong EQ may be expected soon.

**4. Conclusion**

The results presented in this work advocate for the already expressed [11] view that SSB of critical phenomena is of key importance in understanding the creation of a seismic event. The main objective of this work was to extend our study to what happens beyond the completion of the SSB, during the broken symmetry phase, in the time period close-after the occurrence of a strong EQ. The presented findings provide an answer to the question expressed in the title of the present article, that is, whether post-SSB power-laws in the MHz EME after a very strong EQ can be considered an indication for the preparation of a new strong EQ or not.

In the application of MCF to MHz EME after a very strong EQ, beyond the identification of power-laws in the laminar lengths' distribution, we additionally introduced the calculation of the autocorrelation function of the corresponding MHz EME time series excerpt as a necessary criterion for the identification of "true" critical state. Particularly, we discern two possible cases: (a) If a post-SSB power-law in the laminar lengths' distribution is accompanied by a long-range correlation, as proven by means of the autocorrelation function, then the corresponding MHz EME time series excerpt has indeed embedded critical dynamics, it is a "true" "critical window" indicating the first stage in the preparation of a new strong EQ, specifically signalizes the ''critical epoch'', when the short-range correlations evolve into long-range ones [10]. (b) If a post-SSB power-law in the laminar lengths' distribution is not accompanied by a long-range correlation, as proven by means of the autocorrelation function, then the corresponding MHz EME time series excerpt embed "locally surviving" critical dynamics close-after the completion of the SSB and the elimination of the unstable critical point. Such power-laws can be interpreted in analogy to the phase of the domain walls of thermal systems and the corresponding MHz EME are probably due to post-EQ local fracture structures of the heterogeneous system close to the fault.

Finally, it should be mentioned that the complexity of fracture phenomena justifies their approach from different points of view. For example, Maslov et al. [24] have formally established the relationship between spatial fractal behavior and long-range temporal correlations for a broad range of critical phenomena. Therefore, an interesting direction for further investigation could be to study post-SSB power-laws from the spatial point of view too, e.g., by designing and conducting an appropriate laboratory experiment.